\documentclass[12pt]{article}
\usepackage{graphicx}
\usepackage[cp1251]{inputenc}
\usepackage{rotating}
\begin{document}
 \noindent {\footnotesize\it Astronomy Letters, 2017, Vol. 43, Issue 3}
 \newcommand{\dif}{\textrm{d}}

 \noindent
 \begin{tabular}{llllllllllllllllllllllllllllllllllllllllllllll}
 & & & & & & & & & & & & & & & & & & & & & & & & & & & & & & & & & & & & & \\\hline\hline
 \end{tabular}

  \vskip 0.5cm
 \centerline{\bf THE GALAXY KINEMATICS FROM OB STARS }
 \centerline{\bf WITH PROPER MOTIONS FROM THE GAIA DR1 CATALOG}
 \bigskip
 \bigskip
  \centerline
 {
 V.V. Bobylev\footnote [1]{e-mail: vbobylev@gao.spb.ru},
 A.T. Bajkova
 }
 \bigskip
 \centerline {\small \it Central (Pulkovo) Astronomical Observatory, Russian Academy of Sciences}
 \bigskip
 \bigskip
 \bigskip
 {
We consider two previously studied samples of OB stars with
different distance scales. The first one consists of massive
spectroscopic binary stars with photometric distances, and the
second one ~--- with the distances determined along the lines of
interstellar calcium. The OB-stars are located at distances up to
7 kpc from the Sun. They are identified with the Gaia DR1 catalog.
It is shown that the use of the proper motions, taken from the
Gaia DR1 catalog, allows to reduce random errors of determination
of the Galactic rotation parameters in comparison with the
previously known ones. From the analysis of 208 OB stars from the
Gaia DR1 catalog with proper motions and parallaxes with relative
errors less than 200\%  were found the following values of kinematic
parameters: $(U,V)_\odot=(8.67,6.63)\pm(0.88,0.98)$~km s$^{-1}$,
$\Omega_0=27.35\pm0.77$~km s$^{-1}$ kpc$^{-1}$,
$\Omega^{'}_0=-4.13\pm 0.13$~km s$^{-1}$ kpc$^{-2}$,
$\Omega^{''}_0=0.672\pm 0.070$~km s$^{-1}$ kpc$^{-3}$, the Oort
constants $A=-16.53\pm 0.52$~km s$^{-1}$ kpc$^{-1}$ and $B=
10.82\pm0.93$~km s$^{-1}$ kpc$^{-1}$, linear circular velocity of
the Local Standard of Rest around the Galactic center
$V_0=219\pm8$~km s$^{-1}$ for an adopted value of $R_0=8.0\pm
0.2$~kpc. In addition the Galactic rotation parameters were
obtained from only line-of-sight velocities of the same stars.
From the comparison of these two values of $\Omega^{'}_0$ a
distance scale of the Gaia DR1 catalog was determined as a value
close to unit, namely 0.96. From 238 OB-stars of the united sample
with photometric distances for stars of the first sample and
distances in the calcium scale for stars of the second sample,
line-of-sight velocities and proper motions from the Gaia DR1
catalog, were found the following values of the kinematic
parameters: $(U,V,W)_\odot=(8.19,9.28,8.79)\pm(0.74,0.92,0.74)$~km
s$^{-1}$, $\Omega_0=31.53\pm0.54$~km s$^{-1}$ kpc$^{-1}$,
$\Omega^{'}_0=-4.44\pm0.12$~km s$^{-1}$ kpc$^{-2}$,
$\Omega^{''}_0=0.706\pm0.100$~km s $^{-1}$ kpc$^{-3}$, here Oort
constants: $A=-17.77\pm0.46$~km s$^{-1}$ kpc$^{-1}$, $B=
13.76\pm0.71$~km s$^{-1}$ kpc$^{-1}$ and $V_0=252\pm8$~km
s$^{-1}$. }

\section*{INTRODUCTION}
From the combination of the first year Gaia observational
data(Prusti et al., 2016) with the positions and proper motions of
Tycho-2 (H{\o}g et al., 2000) stars was created the Gaia DR1
catalog. It is referred to as TGAS (Tycho--Gaia Astrometric
Solution, Mihalik, et al., 2015; Brown
 et al., 2016; Lindegren et al., 2016), and contains parallaxes and
 proper motion of about 2 million brightest stars (up to $\sim11.^m5$).

Random errors of data included in the Gaia DR1 catalog are
comparable or less than those given in the HIPPARCOS~(1997) and
Tycho-2 catalogs. Average errors of parallaxes are about 0.3 mas
(milliarcseconds). It means that with distances with errors of
about 10\% it is possible to cover the Solar neighborhood with a
radius only of about 300 pc. Therefore, when studying the
structure and kinematics of the Galaxy at larger distances from
the Sun (3~kpc, and more) currently remains relevant approach
using photometric or other distance scales.

For most of the TGAS stars an average error of proper motion is
about 1 mas/yr (milliarcseconds per year). But for a significant
number ($\sim$94000) of the HIPPARCOS catalog stars this error is
an order of magnitude smaller and is about 0.06 mas/yr (Brown et
al., 2016). Therefore, the analysis of their spatial velocities
using high-precision proper motions is of great interest.

In this work two samples of OB stars are used. The first sample
consists of young massive  OB stars with photometric estimates of
distances, most of them are spectroscopic binary stars. This
sample was compiled by Bobylev\&Bajkova~(2013; 2015) on published
sources, and it contains about 300  OB stars. The second sample
basically consists of single OB stars,  with distances defined by
the spectral lines of interstellar calcium. These distances are
determined by Megier et al. (2005; 2009) and Galazutdinov et al.
(2015). The study of this sample with star proper motions from the
HIPPARCOS catalog was made by Bobylev\&Bajkova~(2011; 2015).

The aim of this work is testing of the distance scale of the Gaia
DR1 catalog, re-determining the Galactic rotation parameters from
data on OB stars with proper motions from the Gaia DR1 catalog. We
analyze full spatial velocities of OB stars as well as the
velocities determined only from proper motions or only from line-of-sight velocities.

 \section*{METHODS}\label{method}
From observations we know three components of the star velocity:
the line-of-sight velocity $V_r$ and the two projections of the
tangential velocity $V_l=4.74 r\mu_l\cos b$ and $V_b=4.74 r\mu_b,$
directed along the Galactic longitude $l$ and latitude $b$
respectively and expressed in km s$^{-1}$. Here the coefficient
4.74 is the ratio of the number of kilometers in astronomical unit
by the number of seconds in a tropical year, and $r$ is a
heliocentric distance of the star in kpc. The components of a
proper motion of $\mu_l\cos b$ and $\mu_b$ are expressed in the
mas yr$^{-1}$. The velocities $U,V,W$ directed along the
rectangular Galactic coordinate axes are calculated via the
components of $V_r, V_l, V_b$:
 \begin{equation}
 \begin{array}{lll}
 U=V_r\cos l\cos b-V_l\sin l-V_b\cos l\sin b,\\
 V=V_r\sin l\cos b+V_l\cos l-V_b\sin l\sin b,\\
 W=V_r\sin b                +V_b\cos b,
 \label{UVW}
 \end{array}
 \end{equation}
where $U$ is directed from the Sun to the Galactic center, $V$ is
in the direction of Galactic rotation, and  $W$ is directed toward
the north Galactic pole. We can find two velocities, $V_R$
directed radially away from the Galactic center and $V_{circ}$
orthogonal to it and pointing in the direction of Galactic
rotation, based on the following relations:
 \begin{equation}
 \begin{array}{lll}
  V_{circ}= U\sin \theta+(V_0+V)\cos \theta, \\
       V_R=-U\cos \theta+(V_0+V)\sin \theta,
 \label{VRVT}
 \end{array}
 \end{equation}
where the position angle $\theta$ satisfies to the ratio
$\tan\theta=y/(R_0-x)$, and $x,y,z$ are rectangular
heliocentric coordinates of the star (velocities $U,V,W$ are directed along the axes
$x,y,z$ respectively).

To determine the parameters of the Galactic rotation curve, we
use the equation derived from Bottlinger's formulas in which
the angular velocity $\Omega$ was expanded in a series to terms of
the second order of smallness in $r/R_0 $:
\begin{equation}
 \begin{array}{lll}
 V_r=-U_\odot\cos b\cos l-V_\odot\cos b\sin l\\
 -W_\odot\sin b+R_0(R-R_0)\sin l\cos b\Omega^{'}_0
 +0.5R_0(R-R_0)^2\sin l\cos b\Omega^{''}_0,
 \label{EQ-1}
 \end{array}
 \end{equation}
 \begin{equation}
 \begin{array}{lll}
 V_l= U_\odot\sin l-V_\odot\cos l-r\Omega_0\cos b\\
 +(R-R_0)(R_0\cos l-r\cos b)\Omega^{'}_0
 +0.5(R-R_0)^2(R_0\cos l-r\cos b)\Omega^{''}_0,
 \label{EQ-2}
 \end{array}
 \end{equation}
 \begin{equation}
 \begin{array}{lll}
 V_b=U_\odot\cos l\sin b + V_\odot\sin l \sin b\\
 -W_\odot\cos b-R_0(R-R_0)\sin l\sin b\Omega^{'}_0
    -0.5R_0(R-R_0)^2\sin l\sin b\Omega^{''}_0,
 \label{EQ-3}
 \end{array}
 \end{equation}
where $R$ is the distance from the star to the Galactic rotation axis,
  \begin{equation}
 R^2=r^2\cos^2 b-2R_0 r\cos b\cos l+R^2_0.
 \end{equation}
$\Omega_0$ is the angular velocity of Galactic rotation at the
solar distance $R_0$, the parameters $\Omega^{'}_0$ and
$\Omega^{''}$ are the corresponding derivatives of the angular
velocity, and $V_0=|R_0\Omega_0|$.

It is necessary to adopt a certain value of the distance $R_0$.
One of the most reliable estimates of this value
$R_0=8.28\pm0.29$~kpc obtained by Gillessen et al. (2009) from the
analysis of the orbits of the stars, moving around the
supermassive black hole at the center of the Galaxy. From masers
with trigonometric parallaxes Reid et al. (2014) found
$R_0=8.34\pm0.16$~kpc. From analysis of kinematics of the masers
Bobylev\&Bajkova (2014) estimated $R_0=8.3\pm0.2$~kpc, in work by
Bajkova\&Bobylev (2015) it is found $R_0=8.03\pm0.12$~kpc,
Rastorguev et al.(2016) obtained $R_0=8.40\pm0.12$~kpc. Recent
analysis of the orbits of stars moving around the supermassive
black hole at the center of the Galaxy gave an estimate
$R_0=7.86\pm0.2$~kpc (Boehle et al., 2016). In the present work
the adopted value $R_0=8.0\pm0.2$~kpc.

The influence of the spiral density wave in the radial, $V_R,$ and
residual tangential, $\Delta V_{circ},$ velocities is periodic
with an amplitude of $\sim$10 km s$^{-1}.$ According to the linear
theory of density waves (Lin and Shu 1964), it is described by the
following relations:
 \begin{equation}
 \begin{array}{lll}
       V_R =-f_R \cos \chi,\\
 \Delta V_{circ}= f_\theta \sin\chi,
 \label{DelVRot}
 \end{array}
 \end{equation}
where
 \begin{equation}
 \chi=m[\cot(i)\ln(R/R_0)-\theta]+\chi_\odot
 \end{equation}
is the phase of the spiral density wave ($m$ is the number of
spiral arms, $i$ is the pitch angle of the spiral pattern,
$\chi_\odot$ is the radial phase of the Sun in the spiral density
wave); $f_R$ and $f_\theta$ are the radial and tangential velocity
perturbation amplitudes, which are assumed to be positive.

We apply a spectral analysis to study the
periodicities in the velocities $V_R$ and $\Delta V_{circ}$. The
wavelength $\lambda$ (the distance between adjacent spiral arm
segments measured along the radial direction) is calculated from
the relation
\begin{equation}
 \frac{2\pi R_0}{\lambda}=m\cot(i).
 \label{a-04}
\end{equation}
Let there be a series of measured velocities $V_{R_n}$ (these can
be both radial, $V_R$ and residual tangential, $\Delta
V_{\theta},$ velocities), $n=1,\dots,N$, where $N$ is the number
of objects. The objective of our spectral analysis is to extract a
periodicity from the data series in accordance with the adopted
model describing a spiral density wave with parameters
$f_R,\lambda (i)$ and $\chi_\odot$.

Having taken into account the logarithmic character of the spiral
density wave and the position angles of the objects $\theta_n,$
our spectral (periodogram) analysis of the series of velocity
perturbations is reduced to calculating the square of the
amplitude (power spectrum) of the standard Fourier transform
(Bajkova\&Bobylev 2012):
\begin{equation}
 \bar{V}_{\lambda_k} = \frac{1} {N}\sum_{n=1}^{N} V^{'}_n(R^{'}_n)
 \exp\Bigl(-j\frac {2\pi R^{'}_n}{\lambda_k}\Bigr),
 \label{29}
\end{equation}
where $\bar{V}_{\lambda_k}$ is the $k$th harmonic of the Fourier
transform with wavelength  $\lambda_k=D/k, D$ is the period of the
series being analyzed,
 \begin{equation}
 \begin{array}{lll}
 R^{'}_{n}=R_{\circ}\ln(R_n/R_{\circ}),\\
 V^{'}_n(R^{'}_n)=V_n(R^{'}_n)\times\exp(jm\theta_n).
 \label{21}
 \end{array}
\end{equation}
The algorithm of searching for periodicities modified to properly
determine not only the wavelength but also the amplitude of the
perturbations is described in detail in Bajkova and Bobylev
(2012).

The peak value of the power spectrum $S_{peak}$ corresponds to
the sought-for wavelength $\lambda$. The pitch angle of the spiral
density wave is found from the expression (\ref{a-04}). Amplitude and phase
of perturbations we obtain by the fit of the harmonic with found
wavelength to the measured data. To evaluate the amplitude of perturbations
the following relation can also be used:
 \begin{equation}
f_R(f_\theta)=\sqrt{4\times S_{peak}}.
 \label{Speak}
 \end{equation}
So, our approach consists of two stages: a)constructing a smooth Galactic
rotation curve, and b)spectral analysis of radial
$V_R,$ and residual tangential $\Delta V_{circ}$ velocities. Similar
approach has been applied by Bobylev et al. (2008) for studying the
kinematics of young Galactic objects, by Bobylev\&Bajkova
(2012) for analysis of the Cepheids, and by Bobylev\&Bajkova (2013; 2015)
to determine the Galactic rotation curve from data on massive OB-stars.

 \section*{DATA}\label{data}
In the present work we use the following two samples of OB stars:
1) the sample of spectroscopic binary OB stars and
2) the sample of OB stars with distances determined by the spectral
lines of the interstellar Ca\,II.

The catalog of massive spectroscopic binary stars is described by
Bobylev\&Bajkova~(2013; 2015). There we considered the systems
with the spectra of the main component no later than B2.5 and
different super giants with luminosity classes Ia and Iab. In
addition, from the HIPPARCOS catalog  we took all the B2.5-stars
with relative errors of parallaxes less than 10\%. For all stars
in this catalog, there are estimates of distances, line-of sight
velocities and proper motions. In the end, were collected data on
120 spectroscopic binary and single stars that do not have
properties of ``runaway'' stars, because their residual velocities
do not exceed 40--50~km s$^{-1}$. We identified 98 stars from this
catalog with the stars of Gaia DR1 catalog.

The spectroscopic method of determining star distances using
broadening of the absorption lines of the interstellar ionized
atoms Ca\,II, Na\,I, or K\,I is well known. Megier et al.~(2005;
2009) bound the scale of the equivalent widths to trigonometric
parallaxes of the HIPPARCOS catalogue. It turned out that the
Ca\,II spectral line widths are measured with the most high
accuracy, therefore we call this scale as a calcium scale.
According to the estimates of Megier et al.~(2009), the average
accuracy of individual distance to OB stars is approximately 15\%.

From the analysis of several kinematic parameters obtained from
data on distant OB stars of this sample, Bobylev\&Bajkova
(2011) showed the need for small, no more than 20\%, reducing of
the scale of distances to stars which are further 0.8~kpc from the
Sun. Accounting for the measurements of Galazutdinov et al.
(2015), the whole sample contains 340 OB stars, 168 stars of which
were identified with the stars of the Gaia DR1 catalog.

\begin{figure}[t]
{\begin{center}
   \includegraphics[width=0.6\textwidth]{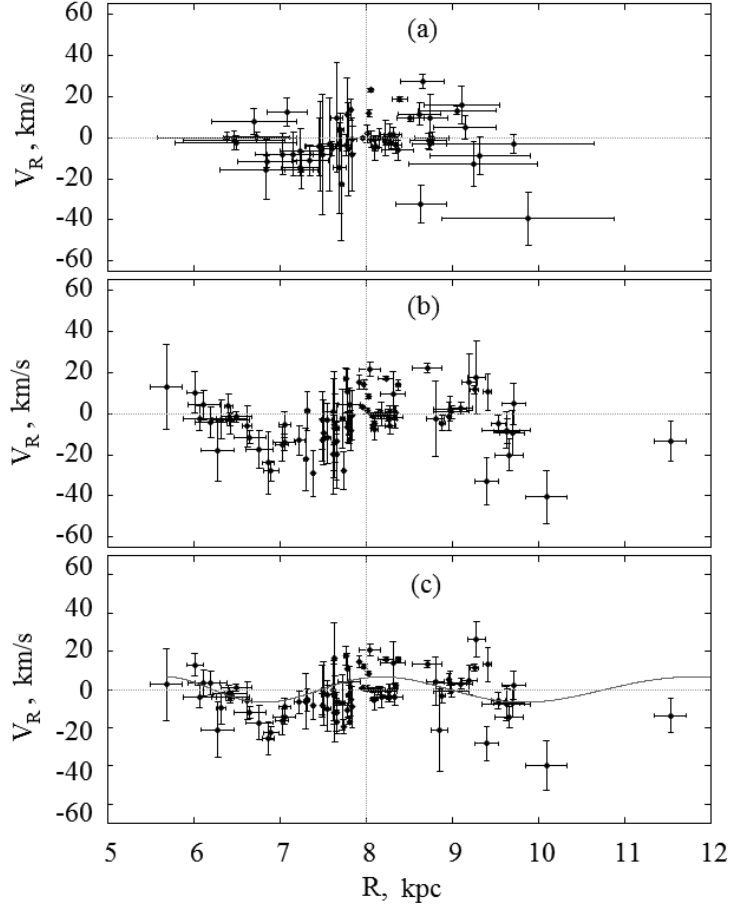}
 \caption{
Radial velocities $V_R$ versus $R$ of
spectroscopic binary OB stars: a)~with distances with relative
errors less than 60\% and proper motions from the Gaia DR1 catalog,
b)~ with photometric distances and proper motions
from the HIPPARCOS catalog, c)~with photometric
distances and proper motions from the Gaia DR1 catalog.
  } \label{f-VR-SB}
\end{center}}
\end{figure}
\begin{figure}[t]
{\begin{center}
   \includegraphics[width=0.4\textwidth]{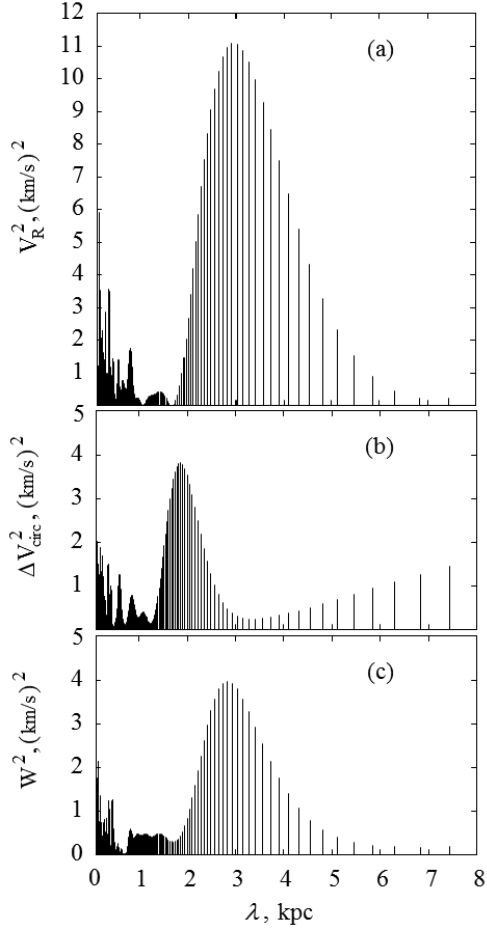}
 \caption{
Power spectra for the spectroscopic binary OB stars with
photometric distances and proper motions from the Gaia DR1
catalog: a)~of the radial velocities $V_R$, b)~of the residual
circular rotation velocities $\Delta V_{circ}$~b) and c)~of the
vertical $W$ velocities.
  } \label{f-SPECTR-SB}
\end{center}}
\end{figure}
\begin{figure}[t]
{\begin{center}
   \includegraphics[width=0.4\textwidth]{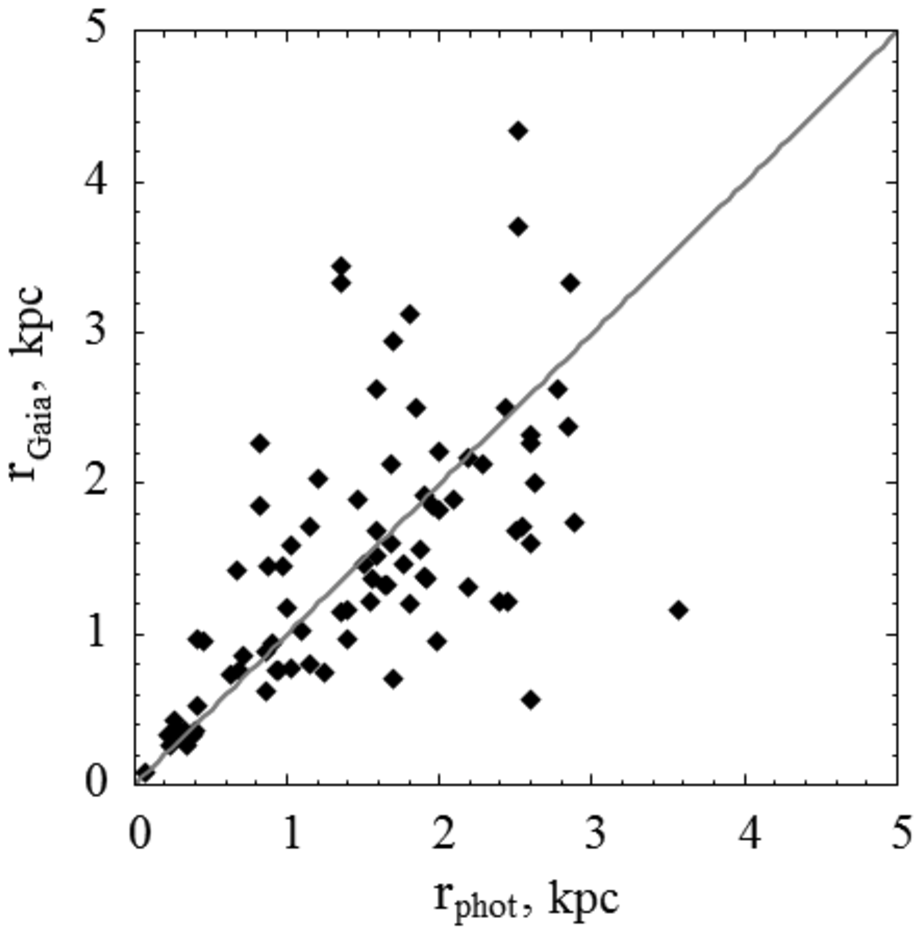}
 \caption{
The distances to the stars from the Gaia DR1 catalog versus
distances to the same stars, determined by the photometric method;
a straight line corresponds to the correlation equal to 1.
  } \label{f-r-r-200}
\end{center}}
\end{figure}

 \section*{RESULTS AND DISCUSSION}\label{results}
The system of conditional equations~(\ref{EQ-1})--(\ref{EQ-3}) was
solved by the least squares method with the following weights
 $w_r=S_0/\sqrt {S_0^2+\sigma^2_{V_r}},$
 $w_l=S_0/\sqrt {S_0^2+\sigma^2_{V_l}}$ and
 $w_b=S_0/\sqrt {S_0^2+\sigma^2_{V_b}},$
where $S_0$ is a ``cosmic'' dispersion, $\sigma_{V_r},
\sigma_{V_l}, \sigma_{V_b}$~ are appropriate error variances of
the observational velocities. The value of $S_0$ is comparable to
the mean square residual $\sigma_0$ (error of unit weight) in the
solution of conditional equations (\ref{EQ-1})--(\ref{EQ-3}). In
the present work $S_0=8$~km s$^{-1}$ is adopted. For the exclusion
of runaway stars, we use the following restriction:
\begin{equation}
 \begin{array}{lll}
  \sqrt{U^2+V^2+W^2}<80~\hbox {km s$^{-1}$},
 \label{criteryi-80}
 \end{array}
\end{equation}
where the velocities $U,V,W$ are residual, they are exempt from
differential rotation of the Galaxy using found pre-rotation
parameters (or already known). Solutions are searched in two
iterations with the exception of large residuals according to the
criterion $3\sigma$.

 \subsection*{Sample of spectroscopic binary stars}\label{results-SB}
In the first step for 98 spectroscopic binary stars subject to
the criterion of~(\ref{criteryi-80}) the original
photometric distances and proper motions were taken from the
HIPPARCOS and Tycho-2 catalogs. From these data were found the following
kinematic parameters:
 \begin{equation}
 \label{solution-1}
 \begin{array}{lll}
 (U_\odot,V_\odot,W_\odot)=(2.7,9.2,10.4)\pm(1.0,1.2,1.1)~\hbox{km s$^{-1}$},\\
      \Omega_0=~28.95\pm0.84~\hbox{km s$^{-1}$ kpc$^{-1}$},\\
  \Omega^{'}_0=-4.15\pm0.17~\hbox{km s$^{-1}$ kpc$^{-2}$},\\
 \Omega^{''}_0=~1.019\pm0.155~\hbox{km s$^{-1}$ kpc$^{-3}$}.
 \end{array}
 \end{equation}
In this solution, the error of unit weight is
$\sigma_0=9.7$~km s$^{-1}$, the Oort constants:
$A=-16.61\pm0.69$~km s$^{-1}$ kpc$^{-1}$ and
$B= 12.35\pm1.08$~km s$^{-1}$ kpc$^{-1}$.
The number of equations was 268.

The solution~(\ref{solution-1}) should be compared with the
obtained by Bobylev\&Bajkova (2015) from data on 120 OB stars
of this sample:
 $(U_\odot,V_\odot,W_\odot)=(2.8,9.2,9.3)\pm(1.0,1.2,1.0)$~km s$^{-1}$,
     $\Omega_0=29.3\pm0.8$~km s$^{-1}$ kpc$^{-1}$,
  $\Omega^{'}_0=-4.28\pm0.15$~km s$^{-1}$ kpc$^{-2}$,
 $\Omega^{''}_0=0.957\pm0.128$~km s$^{-1}$ kpc$^{-3}$,
where the value of the error of unit weight is $\sigma_0=10.5$~km s$^{-1}$,
the linear velocity of the Galactic rotation $V_0=234\pm14$~km s$^{-1}$ (for
$R_0=8.0\pm0.4$~kpc), and the Oort constants:
$A=-17.1\pm0.6$~km  s$^{-1}$ kpc$^{-1}$ and $B=12.2\pm1.0$~km s$^{-1}$ kpc$^{-1}$.

In the second step for these 98 stars the original
photometric distances and proper motions from the Gaia DR1 catalog
were taken. From these data we found the following kinematic
parameters:
 \begin{equation}
 \label{solution-2}
 \begin{array}{lll}
 (U_\odot,V_\odot,W_\odot)=(3.8,8.5,8.4)\pm(0.9,1.1,0.9)~\hbox{km s$^{-1}$},\\
      \Omega_0=~30.91\pm0.74~\hbox{km s$^{-1}$ kpc$^{-1}$},\\
  \Omega^{'}_0=-4.27\pm0.15~\hbox{km s$^{-1}$ kpc$^{-2}$},\\
 \Omega^{''}_0=~0.911\pm0.130~\hbox{km s$^{-1}$ kpc$^{-3}$}.
 \end{array}
 \end{equation}
In this solution, the error of unit weight is
$\sigma_0=8.3$~km s$^{-1}$, the Oort constants:
$A=-17.08\pm0.59$~km s$^{-1}$ kpc$^{-1}$ and
$B= 13.83\pm0.95$~km s$^{-1}$ kpc$^{-1}$.
The number of equations was 265.

In Fig.~\ref{f-VR-SB} the radial velocities $V_R$ versus $R$ are shown
for three cases. The first case represents 60
spectroscopic binary stars with distances with relative errors
less than 60\% and the proper motions taken from the Gaia DR1 catalog.
In the second case the stars with
the photometric distances and the proper motions from
the HIPPARCOS, Tycho-2 and UCAC4 catalogs were used, as in
the solution (\ref{solution-1}). In the third case
the spectroscopic binary OB stars with the photometric distances and
the proper motions from the Gaia DR1 catalog were used, as in
the solution (\ref{solution-2}).

Based on the parameters of the rotation curve represented by the
solution (\ref{solution-2}), we obtained the residual tangential
velocities $\Delta V_{circ}$. In Fig.~\ref{f-SPECTR-SB} the power
spectra of the radial velocities $V_R$, the residual tangential
velocities $\Delta V_{circ}$ and the vertical $W$ ones are shown.
From figure you can see that the power spectrum of the radial
velocities has the largest amplitude.

The parameters of the galactic spiral density wave were obtained
using periodogram analysis of series of the residual tangential
$\Delta V_{circ},$ radial $V_R$ and vertical $W$ velocities. The
amplitudes of the tangential, radial and vertical velocity
perturbations
 $f_\theta=3.9\pm1.6$~km s$^{-1}$,
 $     f_R=6.7\pm1.2$~km s$^{-1}$,
 $     f_W=4.0\pm1.1$~km s$^{-1}$ respectively,
wavelength of the perturbations
 $\lambda_\theta=1.9\pm0.6$~kpc,
 $     \lambda_R=2.9\pm0.4$~kpc and
 $     \lambda_W=2.8\pm0.4$~kpc
for adopted four-armed spiral pattern model ($m=4$).
The Sun's phase in the spiral density wave
 $(\chi_\odot)_\theta=-132^\circ\pm14^\circ$ from the residual tangential velocities,
 $     (\chi_\odot)_R=-157^\circ\pm10^\circ$ from the radial velocities and
 $     (\chi_\odot)_W=-159^\circ\pm11^\circ$ from the vertical velocities.
The bottom panel of Fig.~\ref{f-VR-SB} shows the wave in the radial
velocities, constructed from the parameters found above.

You can see that the parameters of the velocity perturbations found in the residual
tangential velocities are determined with large relative
errors, i.e. at these velocities is it difficult to see the direct influence of the spiral
density wave, perhaps there is a more complicated picture than
we have modelled. On the contrary, the parameters of the velocity perturbation,
found in the radial and vertical velocities are
a in good agreement with each other.

The values of the parameters of the spiral density wave found in
the radial velocities $V_R$, are in good agreement with the
estimated ones obtained for 120 stars of this sample by Bobylev\&Bajkova (2015).
However, in this work we see a good agreement
between the parameters found in the radial and vertical velocities
(except for amplitudes), while in the work of Bobylev\&Bajkova
(2015), it was mentioned a poor agreement between the estimates of
these parameters.

Based on the 87 OB-stars with relative errors of distances from
the Gaia DR1 catalog, not exceeding 200\%, there was built
Fig.~\ref{f-r-r-200}. On it the distances to the stars from the
Gaia DR1 catalog versus the distances to these stars determined by
photometric method are given. We can see that there is practically
no significant systematic differences between these two scales of
distances. When the relative errors of distances from the Gaia DR1
catalog are more than 200\%, the points in the figure strongly
deviate upward from the straight line. Therefore, when such large
errors the distances from the Gaia DR1 catalog are undesirable to
use.

 \subsection*{Sample of OB stars with the calcium distance scale}\label{results-CaII}
In the first step, for 160 stars of this sample that satisfy the
criterion (\ref{criteryi-80}), the distances in the calcium scale
and the proper motions from the HIPPARCOS and Tycho-2 catalogs
were used. The distances of  the stars were multiplied by a
scaling factor 0.8, except those which are close to the Sun
($r<0.8$~kpc).

Using these stars we found the following kinematic parameters:
 \begin{equation}
 \label{solution-1-CaII}
 \begin{array}{lll}
 (U_\odot,V_\odot,W_\odot)=(10.0,9.9,9.4)\pm(1.2,1.6,1.2)~\hbox{km s$^{-1}$},\\
 \Omega_0=~29.15\pm0.93~\hbox{km s$^{-1}$ kpc$^{-1}$},\\
 \Omega^{'}_0=-4.35\pm0.19~\hbox{km s$^{-1}$ kpc$^{-2}$},\\
 \Omega^{''}_0=~0.874\pm0.153~\hbox{km s$^{-1}$ kpc$^{-3}$}.
 \end{array}
 \end{equation}
In this solution, the error of unit weight is $\sigma_0=14.7$~km
s$^{-1}$, the Oort constants: $A=-17.40\pm0.74$~km s$^{-1}$
kpc$^{-1}$ and $B= 11.75\pm1.18$~km s$^{-1}$ kpc$^{-1}$. The
number of equations was 448.

In the second step, for these 160 stars with the same distances,
the proper motions were taken from the Gaia DR1 catalog. From these data
the following kinematics parameters were found:
 \begin{equation}
 \label{solution-2-CaII}
 \begin{array}{lll}
 (U_\odot,V_\odot,W_\odot)=(10.7,9.1,9.0)\pm(1.0,1.3,1.0)~\hbox{km s$^{-1}$},\\
      \Omega_0=~31.17\pm0.70~\hbox{km s$^{-1}$ kpc$^{-1}$},\\
  \Omega^{'}_0=-4.54\pm0.15~\hbox{km s$^{-1}$ kpc$^{-2}$},\\
 \Omega^{''}_0=~0.788\pm0.123~\hbox{km s$^{-1}$ kpc$^{-3}$}.
 \end{array}
 \end{equation}
In this solution, the error of unit weight is
$\sigma_0=12.3$~km s$^{-1}$, the Oort constants:
$A=-18.14\pm0.59$~km s$^{-1}$ kpc$^{-1}$ and
$B= 13.03\pm0.91$~km s$^{-1}$ kpc$^{-1}$.
The number of equations was 465.

We can see that the use of
proper motions, taken from the Gaia DR1 catalog for
the analysis of the stars of both the first and the second samples, reduces
random errors of estimation of the Galactic rotation parameters at
comparison with the previously known results.

 \subsection*{OB stars with data from the Gaia DR1 catalog}\label{results-GaiaDR1}
In the Gaia DR1 catalog there is no information allowing
accurate spectral classification of stars. However, in the present work
we have two samples of stars with known spectral
classification. Based on this material
the combined sample of 238 OB stars has been formed.

To study the distance scale and proper motions of the Gaia DR1
catalog it is interesting to determine the kinematic parameters
using only the velocities $V_l.$ For this the system of
conditional equations of the form~(\ref{EQ-2}) with five unknowns
was solved using the least squares method. The $W$ component of
the Sun velocity has been fixed as $W_\odot=7$~km s$^{-1}$,
because it can not be determined only from equation~(\ref{EQ-2}).

Components $V_l$ have been calculated using parallaxes and proper
motions from the Gaia DR1 catalog. A total number of OB stars that
meet the criterion~(\ref{criteryi-80}), and have parallaxes with
relative errors less than 200\%, is equal to 222. From these data
we found the following kinematic parameters:
 \begin{equation}
 \label{solution-999}
 \begin{array}{lll}
 (U_\odot,V_\odot)=(8.67,6.63)\pm(0.88,0.98)~\hbox{km s$^{-1}$},\\
      \Omega_0=~27.35\pm0.77~\hbox{km s$^{-1}$ kpc$^{-1}$},\\
  \Omega^{'}_0=-4.13\pm0.13~\hbox{km s$^{-1}$ kpc$^{-2}$},\\
 \Omega^{''}_0=~0.672\pm0.070~\hbox{km s$^{-1}$ kpc$^{-3}$}.
 \end{array}
 \end{equation}
In this solution, the error of unit weight is
$\sigma_0=8.0$~km s$^{-1}$, the Oort constants:
$A=-16.53\pm0.52$~km s$^{-1}$ kpc$^{-1}$ and
$B= 10.82\pm0.93$~km s$^{-1}$ kpc$^{-1}$.
The number of equations was 208.

For studying the distance scale of the Gaia DR1 catalog it is
important to compare obtained kinematic parameters with the parameters obtained
using only the line-of-sight velocities of the same stars. In this case only one
equation of type~(\ref{EQ-1}) with four unknowns was solved. For
this case we found the following kinematic parameters:
 \begin{equation}
 \label{solution-888}
 \begin{array}{lll}
 (U_\odot,V_\odot)=(6.9,10.9)\pm(1.7,1.7)~\hbox{km s$^{-1}$},\\
  \Omega^{'}_0=-4.31\pm0.26~\hbox{km s$^{-1}$ kpc$^{-2}$},\\
 \Omega^{''}_0=~0.979\pm0.180~\hbox{km s$^{-1}$ kpc$^{-3}$}.
 \end{array}
 \end{equation}
The number of equations was 217. The error of unit weight is
$\sigma_0=15.6$~km s$^{-1}$, the Oort constant
$A=-17.3\pm1.0$~km s$^{-1}$ kpc$^{-1}$.

There is the well-known method of determining the correction
factor of the distance scale (Zabolotskikh et al., 2002) from the
comparison of the values of the parameter $\Omega^{'}_0$, one of
which does not depend on star distances when analyzing only
line-of-sight velocities. Then we calculate a ratio of the
$\Omega^{'}_0$ values from solutions (\ref{solution-999}) and
(\ref{solution-888}): $4.13/4.31=0.96$, which indicates that there
is practically no need in correction of the distance scale of the
Gaia DR1 catalog.

Finally, the system of equations~(\ref{EQ-1})--(\ref{EQ-3}) was
solved using data on 238 OB stars (98 stars have photometric
distances and 140 stars have the distances determined in calcium
scale). The proper motions for all used stars were taken from the
Gaia DR1 catalog. The distances of the stars from calcium scale
were multiplied by a scaling factor 0.8, except those stars which
are close to the Sun ($r<0.8$~kpc). As a result we have obtained
the following solution:
 \begin{equation}
 \label{solution-3-common}
 \begin{array}{lll}
 (U_\odot,V_\odot,W_\odot)=(8.19,9.28,8.79)\pm(0.74,0.92,0.74)~\hbox{km s$^{-1}$},\\
      \Omega_0=~31.53\pm0.54~\hbox{km s$^{-1}$ kpc$^{-1}$},\\
  \Omega^{'}_0=-4.44\pm0.12~\hbox{km s$^{-1}$ kpc$^{-2}$},\\
 \Omega^{''}_0=~0.706\pm0.100~\hbox{km s$^{-1}$ kpc$^{-3}$}.
 \end{array}
 \end{equation}
In this solution, the error of unit weight is
$\sigma_0=11.1$~km s$^{-1}$, the Oort constants:
$A=-17.77\pm0.46$~km s$^{-1}$ kpc$^{-1}$ and
$B= 13.76\pm0.71$~km s$^{-1}$ kpc$^{-1}$.
The number of equations was 692.

Among the solutions found in the present work, the
solution~(\ref{solution-3-common}) gives the kinematic parameters
with the least errors. It is interesting to compare these
parameters with, for example, the estimates obtained by Rastorguev
et al. (2016) from data on 130 masers with measured trigonometric
parallaxes:
\begin{equation}
 \label{solution-ras}
 \begin{array}{lll}
 (U_\odot,V_\odot)=(11.40,17.23)\pm(1.33,1.09)~\hbox{km s$^{-1}$},\\
      \Omega_0=~28.93\pm0.53~\hbox{km s$^{-1}$ kpc$^{-1}$},\\
  \Omega^{'}_0=-3.96\pm0.07~\hbox{km s$^{-1}$ kpc$^{-2}$},\\
 \Omega^{''}_0=~0.87\pm0.03~\hbox{km s$^{-1}$ kpc$^{-3}$},\\
 V_0=243\pm10 ~\hbox{km s$^{-1}$}
 \end{array}
 \end{equation}
for found value $R_0=8.40\pm0.12$~kpc.

We can see that the values of the Galactic rotation parameters
found by us ~(\ref{solution-2}),
(\ref{solution-999}), (\ref{solution-2-CaII}) and
(\ref{solution-3-common}) are in good agreement with ones (\ref{solution-ras}),
 but there is difference in the velocities
$U_\odot$ and $V_\odot$.

 \section*{CONCLUSIONS}\label{conclusions}
We considered two well-studied earlier by us samples of OB stars
with the different scales of distances. The first one consists of
massive spectroscopic binary OB stars, the distances to which are
mainly determined by photometrical method, for small part of this
sample there are trigonometric or dynamic parallaxes. The second
sample consists of OB stars with distances determined by lines of
interstellar calcium. All these OB stars are located in wide area
of the Solar neighborhood, with a radius of about 7 kpc, therefore
they are well suited for studying the Galactic rotation. The
identification of these stars with the Gaia DR1 catalog is
fulfilled.

It is shown, that the use of the proper motions from
the Gaia DR1 catalog,reduces random errors of the sought-for
Galactic rotation parameters in comparison with the parameters previously determined.

From the analysis of only proper motions of 208 OB stars from the
Gaia DR1 catalog with a relative error of parallaxes less
than 200\%, the following values of the kinematic parameters were found:
 $(U,V)_\odot=(8.67,6.63)\pm(0.88,0.98)$~km s$^{-1}$,
      $\Omega_0=27.35\pm0.77$~km s$^{-1}$ kpc$^{-1}$,
 $\Omega^{'}_0=-4.13\pm0.13$~km s$^{-1}$ kpc$^{-2}$,
 $\Omega^{''}_0=0.672\pm0.070$~km s$^{-1}$ kpc$^{-3}$,
the values of the Oort constants: $A=-16.53\pm0.52$~km s$^{-1}$
kpc$^{-1}$ and $B=10.82\pm0.93$~km s$^{-1}$ kpc$^{-1}$, circular
linear rotation velocity of the Local Standard of Rest around the
axis of rotation of the Galaxy is equal to $V_0=219\pm8$~km
s$^{-1}$ for an adopted value of $R_0=8.0\pm0.2$~kpc. This
solution  is the most interesting result of this work, because
there were used completely new observational data. The parameters
found show, that despite of large errors of the parallaxes, there
is a good agreement with the values determined using only line-of-sight
velocities of these OB stars. From comparison of the two values of $\Omega^{'}_0$,
the first of which was obtained using only proper motions and the second one --- only line-of-sight
velocities (which does not depend on star distances), we found that the
correction coefficient of the distance scale of the Gaia DR1
catalog is equal to 0.96.

By another approach, for 238 OB stars we used line-of-sight velocities,
the photometric distances for 98 stars of the first sample and
the distances in calcium scale for 140 stars of the second sample, and
the proper motions for all stars from the Gaia DR1 catalog. From these data we obtained
the following values of the kinematic parameters:
 $(U,V,W)_\odot=(8.19,9.28,8.79)\pm(0.74,0.92,0.74)$~km s$^{-1}$,
      $\Omega_0=31.53\pm0.54$~km s$^{-1}$ kpc$^{-1}$,
 $\Omega^{'}_0=-4.44\pm0.12$~km s$^{-1}$ kpc$^{-2}$,
 $\Omega^{''}_0=0.706\pm0.100$~km s$^{-1}$ kpc$^{-3}$,
The Oort constants: $A=-17.77\pm0.46$~km s$^{-1}$ kpc$^{-1}$ and
$B= 13.76\pm0.71$~km s$^{-1}$ kpc$^{-1}$. The linear circular
velocity of the Local Standard of Rest $V_0=252\pm8$~km s$^{-1}$
for adopted $R_0=8.0\pm0.2$~kpc.

In our opinion, the distance scale of the first
sample is the best one. For 98 stars of this sample, using
their photometric distances, line-of-sight velocities, as well as proper
motions from the Gaia DR1 catalog we obtained the solution~(\ref{solution-2}).
The parameters of the galactic spiral density wave
were obtained using spectral (periodogram)
analysis of series of the residual tangential
$\Delta V_{circ},$ radial
$V_R$ and vertical $W$ velocities
of this sample. The amplitudes of the tangential, radial and vertical
velocity perturbations
 $f_\theta=3.9\pm1.6$~km s$^{-1}$,
 $     f_R=6.7\pm1.2$~km s$^{-1}$,
 $     f_W=4.0\pm1.1$~km s$^{-1}$ respectively,
the wavelength of the perturbations
 $\lambda_\theta=1.9\pm0.6$~kpc,
 $     \lambda_R=2.9\pm0.4$~kpc and
 $     \lambda_W=2.8\pm0.4$~kpc
for the accepted four-armed model of the spiral pattern ($m=4$).
Phase of the Sun in the spiral density wave:
 $(\chi_\odot)_\theta=-132^\circ\pm14^\circ$ from residual tangential velocities,
 $     (\chi_\odot)_R=-157^\circ\pm10^\circ$ from radial velocities and
 $     (\chi_\odot)_W=-159^\circ\pm11^\circ$ from vertical velocities.

 \medskip

\subsection*{ACKNOWLEDGEMENTS}
This work was supported by the ``Transient and Explosive Processes in
Astrophysics'' Program P--7 of the Presidium of the Russian
Academy of Sciences.

 \bigskip\medskip{REFERENCES}\medskip{\small

 1.A.T. Bajkova, V.V. Bobylev, Astron. Lett. {\bf 38}, 549 (2012).

 2.A.T. Bajkova, V.V. Bobylev, Baltic Astronomy {\bf 24}, 43 (2015).

 3.V.V. Bobylev, A.T. Bajkova, A.S. Stepanishchev, Astron. Lett., {\bf 34}, 515 (2008).

 4.V.V. Bobylev, A.T. Bajkova,  Astron. Lett. {\bf 37}, 526 (2011).

 5.V.V. Bobylev, A.T. Bajkova, Astron. Lett. {\bf 38}, 638 (2012).

 6.V.V. Bobylev, A.T. Bajkova, Astron. Lett. {\bf 39}, 532 (2013).

 7.V.V. Bobylev, A.T. Bajkova, Astron. Lett. {\bf 40}, 389 (2014).

 8.V.V. Bobylev, A.T. Bajkova, Astron. Lett. {\bf 41}, 473 (2015). 

 9. A. Boehle, A.M. Ghez, R. Schodel, L. Meyer, S. Yelda, S. Albers,
    G.D. Martinez, E.E. Becklin, et al., arXiv: 160705726 (2016).  

 10.A.G.A. Brown, A. Vallenari, T. Prusti,J. de Bruijne, F. Mignard, R. Drimmel, et al., arXiv: 1609.04172, (2016).

 11.G.A. Galazutdinov, A. Strobel, F.A. Musaev, A. Bondar, and J. Kre\l owski, PASP {\bf 127}, 126 (2015).

 12.S. Gillessen, F. Eisenhauer, T.K. Fritz, H. Bartko, K. Dodds-Eden,
     O.~Pfuhl, T. Ott, and R. Genzel, Astroph. J. {\bf 707}, L114 (2009).  

 13.E. H{\o}g, C. Fabricius, V.V. Makarov, U. Bastian, P. Schwekendiek,
     A. Wicenec, S. Urban, T. Corbin, and G. Wycoff,
     Astron. Astrophys. {\bf 355}, L~27 (2000). 

 14.C.C. Lin, F.H. Shu, Astrophys. J. {\bf 140}, 646 (1964).

 15.L. Lindegren, U. Lammers, U. Bastian, J. Hernandez, S. Klioner,
     D. Hobbs, A. Bombrun, D. Michalik, et al., arXiv: 1609.04303 (2016).

 16.A.~Megier, A. Strobel, A. Bondar, F.A. Musaev, I. Han,
     J. Kre\l owski, and G.A. Galazutdinov, Astrophys. J. {\bf 634}, 451 (2005).

 17.A. Megier, A. Strobel, G.A. Galazutdinov, and J. Kre\l owski,
     Astron. Astrophys. {\bf 507}, 833 (2009).

 18.D. Michalik, L. Lindegren, and D. Hobbs,
     Astron. Astrophys. {\bf 574}, A115 (2015).

 19.T. Prusti, J.H.J. de Bruijne, A.G.A. Brown,
     A. Vallenari, C. Babusiaux, C.A.L. Bailer-Jones, U. Bastian, M. Biermann, et al.,
     arXiv: 1609.04153, (2016).

 20.A.S. Rastorguev,  M.V. Zabolotskikh, A.K. Dambis,
     N.D. Utkin, A.T. Bajkova, and V.V. Bobylev, arXiv: 1603.09124 (2016).

 21.M.J. Reid, K.M. Menten, A. Brunthaler, X.W. Zheng, T.M. Dame,
     Y. Xu, Y.~Wu, B. Zhang, et al., Astrophys. J. {\bf 783}, 130 (2014). 

 22. The HIPPARCOS and Tycho Catalogues, ESA SP--1200 (1997).

 23.M.V. Zabolotskikh, et al.,  Astron. Lett. {\bf 28}, 454 (2002).

}
 \end{document}